\documentclass[journal]{aa} 
\usepackage{graphicx}
\usepackage{txfonts}

\usepackage[colorlinks=true, linkcolor=blue, citecolor=blue, urlcolor=blue]{hyperref}

\usepackage{ulem}
\usepackage{xcolor}

\definecolor{skyblue}{HTML}{3BA3D4}

\titlerunning{The faint voice of a radio-weak BL Lacertae}

\begin{document}

   \title{The faint voice of a radio-weak BL Lacertae: modeling the broadband emission of WISE~J141046.00+740511.2}

   \subtitle{}

   \author{A. M. Carulli
          \inst{1,2}
          \and
          F. L. Vieyro\inst{3,4}
           \and
          M. M. Reynoso \inst{1,2}
          \and 
          E. J. Marchesini \inst{5}
          \and
          I. Andruchow \inst{3,4}
          }

   \institute{Instituto de Investigaciones Físicas de Mar del Plata (IFIMAR), CONICET-UNMdP, Funes 3350 (7600), Mar del Plata, Argentina
\\
              \email{amcarulli@mdp.edu.ar}
         \and
Departamento de Física, Facultad de Ciencias Exactas y Naturales, Universidad Nacional de Mar del Plata, Funes 3350 (7600), Mar del Plata, Argentina
         \and
             Facultad de Ciencias Astronómicas y Geofísicas, Universidad Nacional de La Plata, Paseo del Bosque, B1900FWA La Plata,
Argentina
             \and
             Instituto Argentino de Radioastronomía, CONICET-CICPBA-UNLP, CC5 (1894) Villa Elisa, Prov. de Buenos Aires, Argentina
             \and
             INAF – Osservatorio di Astrofisica e Scienza dello Spazio, Via Gobetti 93/3, 40129 Bologna, Italy
}

   \date{Received-; accepted -}

  \abstract
   {The WISE source, J141046.00+740511.2, has been recently observed from radio to $\gamma$ rays. Although the optical spectrum is consistent with a BL Lacertae (BL Lac) object, the source displays unusually weak radio emission, which challenges standard interpretations.}
   {Our aim is to understand the origin of the broadband emission from J141046.00+740511.2, using a leptonic model of an extended jet.}
   {To obtain the distribution of electrons along the conical jet, we solved a steady-state convective transport equation. Emissivities were computed along the jet and integrated over the cone volume to obtain the observed flux.}
   {Our model successfully reproduces the observed multiwavelength spectral energy distribution from radio to $\gamma$ rays and naturally accounts for the source's low radio flux without invoking extra emission zones. We also reproduce the mid-IR emission within the same framework.}
   {These results demonstrate that extended jet leptonic models can robustly describe the broadband physics of radio-weak BL Lacs.}
 
   \keywords{galaxies: active -- 
               BL Lacertae objects: general --
               radiation mechanisms: non-thermal --
               gamma rays: galaxies --
               X-rays: galaxies --
               radio continuum: galaxies
               }

   \maketitle
\nolinenumbers

\section{Introduction}

Active galactic nuclei (AGNs) that launch relativistic jets oriented at small angles to the line of sight are known as blazars. This orientation lies within $10^\circ$ of the line of sight \citep{UrryPadovani1995}, which enhances the observed emission due to relativistic beaming and thus boosts it to very high energies ($E > 100 \ \rm GeV$).
Blazars are typically classified into two groups based on their optical spectra \citep{Stickel1991}. The first group consists of flat spectrum radio quasars (FSRQs). These exhibit broad ($>5 \ $Å) and luminous ($\geq 10^{42} \rm erg \ s^{-1}$) optical emission lines due to fast-moving gas near the central engine \citep{Hovatta2019}. The second group includes BL Lacertae objects (BL Lacs). These exhibit very weak or even absent emission lines \citep{Ghisellini2011}.

The spectral energy distributions (SEDs) of both FSRQs and BL Lacs display two characteristic components: a lower-energy and high-energy component. The lower-energy component peaks between the infrared and X-ray bands and is associated with the synchrotron emission of relativistic electrons. The high-energy component, in contrast, peaks between MeV and TeV energies, and its origin depends on the modeling approach: pure leptonic models attribute it to inverse Compton (IC) scattering, while lepto-hadronic models attribute it to proton synchrotron and hadronic processes \citep{Mucke2001}. In pure leptonic models, IC scattering occurs either with low-energy photons produced by the synchrotron emission of the same electrons—known as synchrotron self-Compton (SSC)—or with photons that constitute an external field (e.g., disk radiation, torus, and broad-line regions) known as external Compton (EC) \citep{Prandini2022}. The former is preferred when modeling BL Lacs, while the latter is commonly used for FSRQs \citep{Hovatta2019}.

Considering the frequency at which the synchrotron component peaks, BL Lacs can be classified into four categories \citep{Padovani1995}. These are low-frequency peaked BL Lac objects (LBLs), with $\nu_{\rm peak} < 10^{14} \ \rm Hz$; intermediate-frequency peaked BL Lac objects (IBLs), with $10^{14} < \nu_{\rm peak} < 10^{15} \ \rm Hz$; high-frequency peaked BL Lac objects (HBLs), with $10^{15} < \nu_{\rm peak} < 10^{17} \ \rm Hz$; and extreme high-frequency peaked BL Lac objects (EHBLs), with $\nu_{\rm peak} > 10^{17} \ \rm Hz$.

Blazars are historically considered to be radio-loud. In particular, BL Lacs commonly show large-scale radio emission typical of Fanaroff–Riley type I (FR I) or type II (FR II) galaxies \citep{Kollgaard1992,Urry1995,Giommi2012}, suggesting that they are their beamed counterparts \citep{Ghisellini1998}. Some BL Lacs (e.g., 3C 273 and 3C 286) are even used as calibrators for radio observatories \citep{Bruni2018}. However, recent campaigns have identified objects that share the same optical features as BL Lacs but are radio-weak \citep{Massaro2017, Bruni2018,Marchesini2023}. These sources are known as radio-weak BL Lacs (RWBLs), although their physical interpretation remains under debate.

In particular, \citet{Marchesini2023} conducted an observational campaign and suggested that the Fermi object WISE J141046.00+740511.2 is an RWBL. The authors also developed a leptonic one-zone model to fit the SED of this source in good agreement with both power-law and log-parabolic distributions and found that the synchrotron peaks above $10^{17} \rm Hz$, placing it in the HBL or EHBL regime according to one-zone models. However, the emission resulting from the one-zone model underestimates the far-IR data and radio emission. This limitation motivates the inclusion of extended regions within the jet: if synchrotron emission is self-absorbed in the one-zone model, the extended jet allows radio photons to escape from regions farther out, where self-absorption is weaker. Moreover, the cumulative emission along the extended jet contributes to the far-IR flux, providing a better match to the observations.

Motivated by these results, we developed a purely leptonic model consisting of a convective conical jet, which we applied to J141046.00+740511.2 with the aim of reproducing its radio-to-far-infrared emission. This article is organized as follows. In Sect. \ref{model} we present the basic framework of the model. In Sect. \ref{distribution} we describe the steady-state convective transport equation. Our estimation of nonthermal emissivities is presented in Sect. \ref{emission}. Our model results are shown in Sect. \ref{results} and discussed in Sect. \ref{discussion}. Our framework successfully reproduces the WISE and Fermi-LAT observations and naturally explains J141046.00+740511.2's radio weakness without invoking additional emission zones.

\section{The model}\label{model}

\begin{figure}[!t]
\centering
\includegraphics[width=0.7\linewidth]{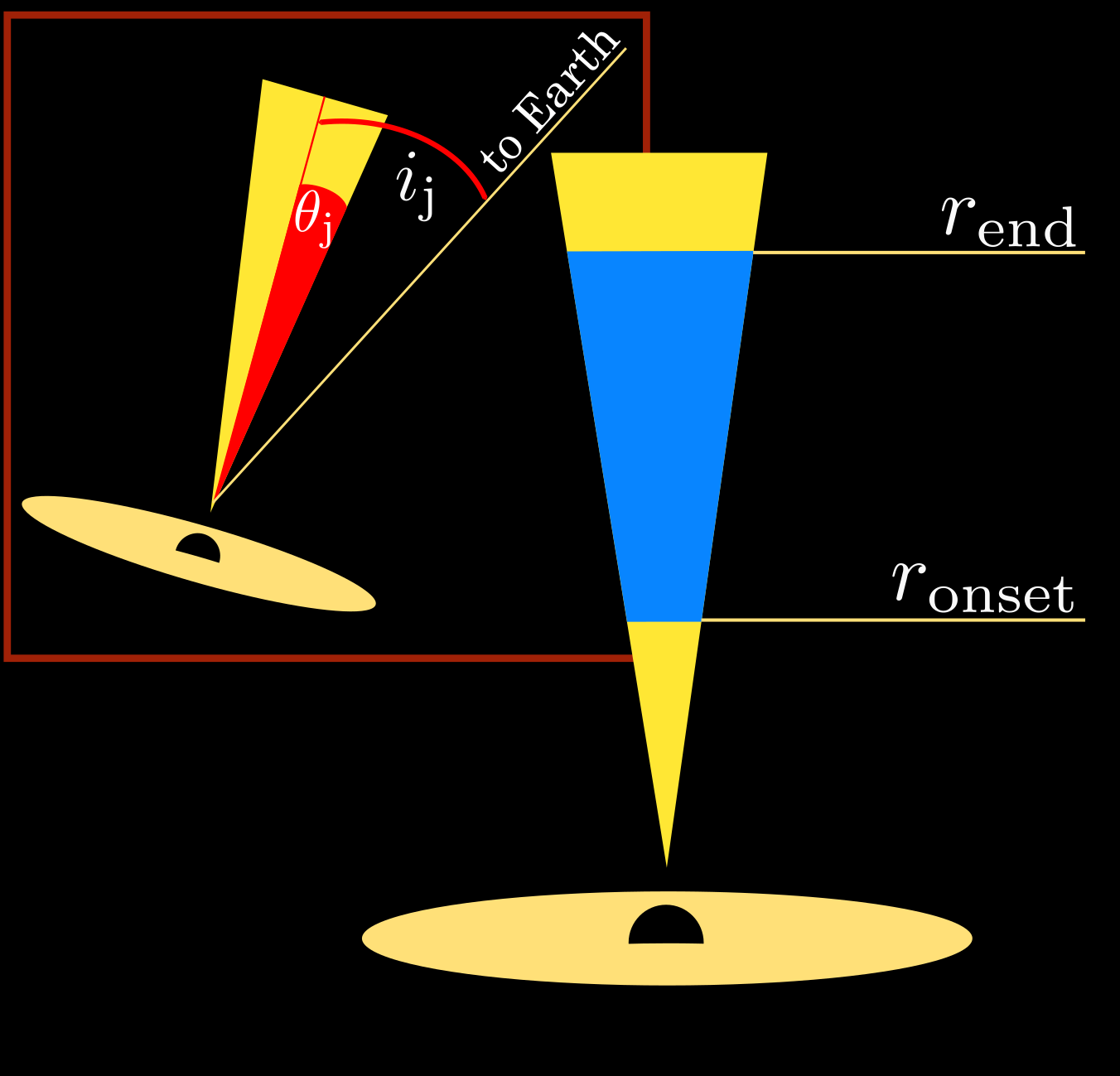}
\caption{Schematic view of the blazar in our model. We show the point of injection ($r_{\rm onset}$) where the convection zone begins and ends ($r_{\rm end}$). We also show the opening angle of the jet $\theta_{\rm j}$ and the angle of the jet with respect to the line of sight, $i_{\rm j}$.}
\label{fig:Figura}
\end{figure}

In this section, we present our adopted model to describe the nonthermal emission of RWBL J141046.00+740511.2. The structure of AGN jets has been examined in several studies, which find that they typically exhibit conical geometries on scales from parsecs up to a few kiloparsecs, as revealed by Very Long Baseline Array (VLBA) observations \citep{Pushkarev2017}. There is also evidence of parabolic shapes in the jet near the central engine \citep{Asada2012, Kovalev2020}. Previous models accounting for a conical region have successfully reproduced BL Lac emission \citep{Potter2012}, as have those adopting a parabolic jet base together with a conical deceleration region \citep{Potter2013,Potter2013b,Potter2013c}. Thus, we consider a conical region of the jet where relativistic electrons are injected at its base and transported convectively along the jet while cooling through synchrotron, SSC, and adiabatic losses. The difference in our approach lies in the way we solve the steady-state convective transport equation along the jet. This allows us to obtain the spatial evolution of the electron distribution self-consistently. The length of the truncated conical region is assumed to be $\sim 100 \rm \ pc$ since the source is not resolved by VLA observations \citep{Marchesini2023}. We show a schematic view of the model in Fig. \ref{fig:Figura}, indicating the position at which the conical region begins, $r_{\rm onset}$, the position at which it ends, $r_{\rm end}$, the opening angle of the jet $\theta_{\rm j}$, and the angle of the jet with respect to the line of sight, $i_{\rm j}$.

We consider that the fluid moves with a constant Lorentz factor $\Gamma$, as measured in the frame attached to the black hole (BH). This means that the plasma moves relative to the BH frame, while the structure of the jet remains fixed. Conversely, in the fluid frame, the plasma remains constant, while the beginning and end positions of the jet vary with time. The motivation behind assuming a constant bulk Lorentz factor along the emitting region is that this zone lies beyond the main acceleration region of the jet, where the flow is expected to be weakly magnetized. We verified that moderate variations in velocity do not significantly affect our results. Finally, in our model we adopted a steady-state but spatially inhomogeneous approach, in which physical quantities vary only along the jet.
In particular, we adopted the following prescription for the magnetic field along the jet. We defined the magnetic field at $r_{\rm onset}$ as
\begin{equation}
B_{\rm onset} = q_{\rm mag} \, \sqrt{8\pi \, \rho_{\rm kin,0}},
\end{equation}
where $\rho_{\rm kin,0}$ is the kinetic energy density of the flow evaluated at $r_{\rm onset}$. This normalization corresponds to a sub-equipartition magnetic field for the adopted parameters.

Beyond this point, it is assumed to decrease along the jet as
\begin{equation}
B(r) = B_{\rm onset} \left( \frac{r_{\rm onset}}{r} \right),
\end{equation}
consistent with magnetic energy conservation along the flow \citep{Potter2012}.

Since the source is not resolved by VLA observations, only upper limits can be placed on the size of the emitting region \citep{Marchesini2023}. For redshifts of $z \sim  0.2$, as adopted in previous work, the emitting region is constrained to be approximately less than or equal to kiloparsec scales. Nevertheless, to remain conservative, we adopted a total extension of $100$ pc. The motivation for assuming a constant Lorentz factor along this region is that the emitting zone is expected to lie beyond the jet's acceleration and collimation region.

\section{Distribution of electrons}\label{distribution}

To obtain the electron distribution along the extended conical jet region, we considered the following transport equation \citep{zdziarski2014}:
\begin{equation} \label{transportEq1}
    \frac{v_{\rm j} \Gamma}{r^2} \frac{\partial(r^2 N_e)}{\partial r} - \frac{\partial(b_e N_e)}{\partial E} = Q_e.
\end{equation}
\noindent Here, jet bulk velocity ($v_{\rm j}$) represents convection, while the second term, $-\partial(b_e N_e)/\partial E$, accounts for continuous energy losses. The variable $r$ is the radius in spherical coordinates, with its origin at the BH. In the following, radial distances $r$ are measured in the laboratory frame. Conversely, particle injections and energies are given in the jet co-moving frame, unless explicitly indicated by the subscript ``L,'' as in $E_{\rm L}$. The term $Q_e$ denotes the electron injection, which we assume to be
\begin{equation}
 \frac{dN_e}{dE d\Omega dV dt}  \equiv   Q_e(r,E) \ = \ k_0 \, E^{-\alpha} \, e^{-E/E_{e, \rm max}} \, \delta(r-r_{\rm onset}),
\end{equation}
\noindent where $r_{\rm onset}$ is the position along the jet where the conical region of the jet begins. The value $E_{\rm max}$ is considered a free parameter in the model, and $k_0$ is obtained by normalizing the injection with electron luminosity. Given that $r_{\rm onset}$ is nonstationary in the co-moving frame, luminosity is computed in the laboratory frame, where the position $r_{\rm onset}$ remains fixed.

Then, electron luminosity is expressed in the laboratory frame attached to the black hole as
\begin{equation}
    L_e \ = \ \int_{E_{\rm L, \ \rm min}}^{\infty} \int_{V} \int_{\Omega} Q_e^{\rm L}(r,E_{\rm L}) \, E_{\rm L} \, d{\Omega} dV dE_{\rm L},
\end{equation}
\noindent where $dV = \ r^2 \sin{\theta}_{\rm L} dr d\theta_{\rm L} d\phi$ in spherical coordinates. Considering relativistic invariants \citep{Dermer2002}, $Q_e^{\rm L}(r,E_{\rm L})$ can be expressed in terms of $Q_e(r,E)$, yielding
\begin{equation}
    Q_e^{\rm L}(r,E_{\rm L}) \ = \ \sqrt{\frac{E_{\rm L}^2-m_e^2 c^4}{E^2-m_e^2 c^4}} Q_e(r,E).
\end{equation}
\noindent Therefore, we compute luminosity via
\begin{equation}
\begin{split}
    L_e \ = \ 4 \pi^2 r_{\rm onset}^2 \ (1-\cos\theta_j) \int_{0}^{\pi} d\theta \sin\theta 
    \int_{E_{\rm L, \  min }}^{\infty} dE_{\rm L} \times \  \  \\ \times k_0 \, E^{-\alpha} \, e^{-E/E_{e, \rm max}} \, E_{\rm L} \, \sqrt{\frac{E_{\rm L}^2-m_e^2 c^4}{E^2-m_e^2 c^4}} ,
    \end{split}
\end{equation} 
\noindent where
\begin{equation}
    E_{\rm L, \ min} \ = \ \frac{E_{\rm min} + \beta \cos\theta \sqrt{\Gamma^2 m_{e}^2 c^4 (\beta^2 \cos^2{\theta} - 1 ) + E_{\rm min}^2 }}{\Gamma (1- \beta^2 \cos^2{\theta})} 
\end{equation}
\noindent and
\begin{equation}
    E \ = \ \Gamma \left( E_{\rm L} - \beta \cos\theta \sqrt{E_{\rm L}^2 - m_{e}^2 c^4} \right) .
\end{equation}

We consider electrons to cool mainly by synchrotron radiation, IC scattering, and adiabatic expansion. In particular, we assume that the dominant target of IC is synchrotron radiation or, more commonly, SSC. Equation (\ref{transportEq1}) can be solved by following the characteristic curve method. This method is useful for converting a partial differential equation (PDE) into a system of ordinary differential equations (ODEs) along curves known as characteristic curves. At the outer boundary of the computational domain, we impose a boundary condition that allows particles to leave the region where the emission is calculated.

\section{Nonthermal emission}\label{emission}

As mentioned above, we considered the SSC process, in which soft photon targets are produced by the same electrons through synchrotron radiation. In Appendix \ref{nt-processes} we include the expressions used to calculate the emissivity of the synchrotron radiation, $Q_{\gamma,\rm Synchr}(r, E_{\gamma})$, including the synchrotron self-absorption coefficient and the procedure for estimating IC emissivity, $Q_{\gamma,\rm IC}(r, E_{\gamma})$.

Once we obtained the emissivity contributions $Q_{\text{ph}}(r, E)$ for each energy $E$ and position $r$, we integrated over the truncated cone volume to get the intensity:
\begin{equation} \label{Qphtot}
\frac{dN_{\rm ph}}{dE_{\rm L} d\Omega dt_{\rm L}} \ = \ \int dV Q_{\rm ph, L}(r,E_{\rm L}),
\end{equation}
\noindent where $Q_{\rm ph, L}(r,E_{\rm L}) \ = \ D \ Q_{\rm ph}(r,E_{\rm L}/D)$.

For an observer at $z=0$, $d\Omega/dA_{\rm obs} = d_M^{-2}$, where $d_M = d_L/(1+z)$ is the transverse co-moving distance. Using $E_{\rm obs} = E_{\rm L}/(1+z)$ and $dt_{\rm obs} = (1+z)\,dt_{\rm L}$, the SED, defined as $E_{\rm obs}^2 F(E_{\rm obs})$ and expressed in units of $\rm erg\,cm^{-2}\,s^{-1}$, is obtained as
\begin{equation}
E_{\rm obs}^2 F(E_{\rm obs}) =
(1+z)^2 \frac{E_{\rm obs}^2}{d_L^2}
\frac{dN_{\rm ph}}{dE_{\rm L}\, d\Omega\, dt_{\rm L}}
\, e^{-\tau_{\rm ext}(z,E_{\rm obs})},
\end{equation}
where $F(E_{\rm obs})$ is the differential photon flux and $\tau_{\rm ext}(z,E_{\rm obs})$ is the optical depth due to extragalactic background light absorption \citep{Franceschini2018}.

To verify our numerical implementation, we compared a one-zone configuration with the public code AM3\footnote{\url{https://am3.readthedocs.io/}} (Astrophysical Multi-Messenger Modeling Software, \citealt{Klinger2024}), finding good agreement across most of the energy range, with minor deviations at the highest energies.

\section{Results}\label{results}

In exploring the parameter space, we varied the intensity of the magnetic field, which depends both on the energy fraction $\epsilon_{\rm B}$ and on jet distance $r_{\rm onset}$, to reproduce the infrared emission. The minimum electron Lorentz factor, $\gamma_{\rm min}$, controls the contribution of the synchrotron spectrum to the radio band: lower or higher values of this parameter either underestimate or overproduce low-frequency emission, respectively. The maximum electron energy, $E_{\rm max}$, is constrained by X-ray data and Fermi-LAT upper limits. A similar effect is observed for the injected spectral index $\alpha$, since steeper values underestimate the X-ray flux, while harder values exceed Fermi-LAT constraints. Considering the impact of these parameters helps reduce the degeneracy in the model.

\begin{figure}[!t]
\centering
\includegraphics[width=1\linewidth]{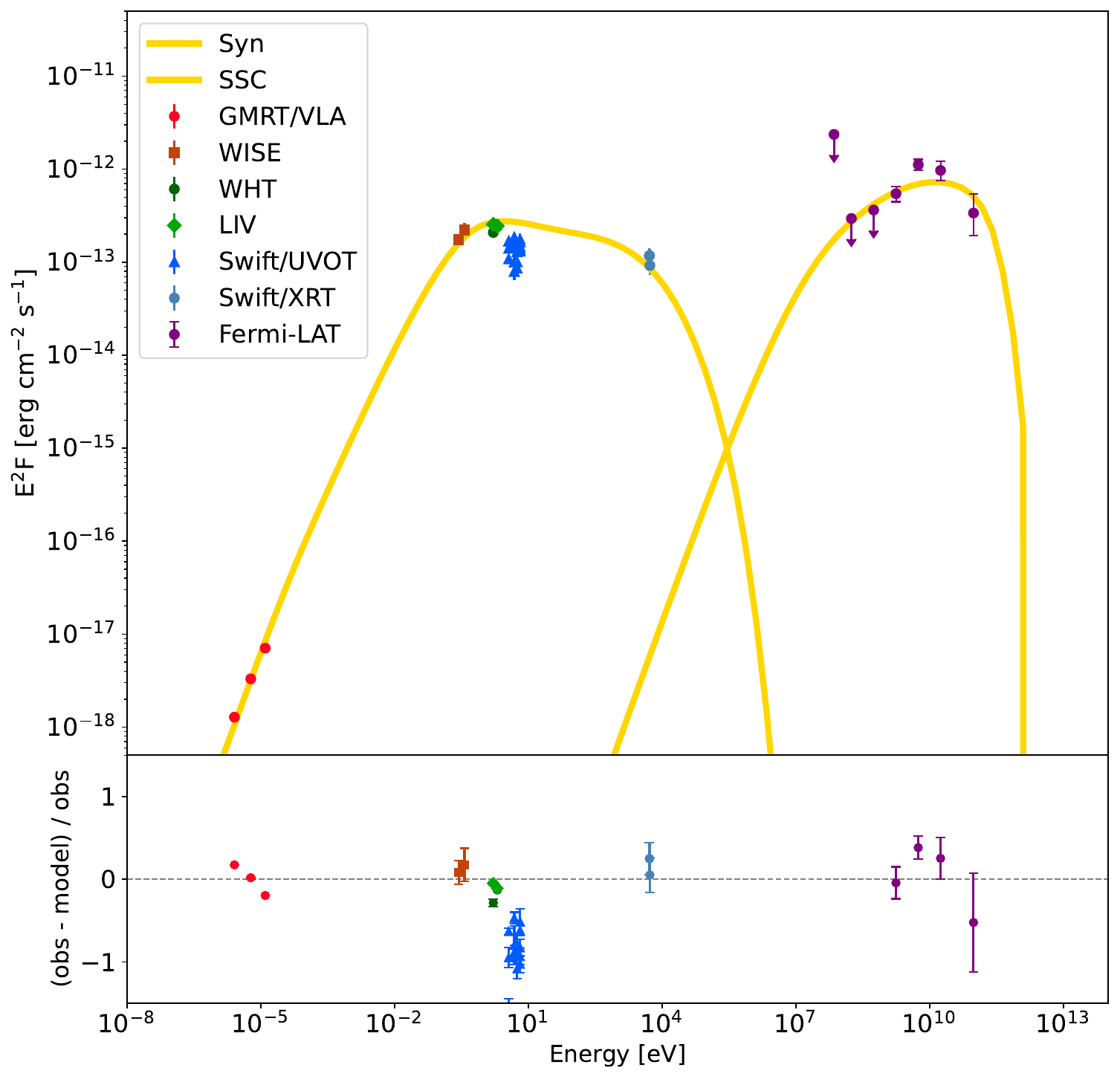}
\caption{Gamma-ray fluxes at Earth produced by synchrotron and SSC radiation corresponding to the parameters in Table \ref{tabla}.}
\label{fig:photonflux}
\end{figure}

In Fig. \ref{fig:photonflux}, we present the resulting photon fluxes using the set of parameters listed in Table \ref{tabla}. The resulting SED exhibits a synchrotron component with a peak frequency of $\sim 5.54\times 10^{14}$ Hz, implying an IBL blazar. The radio-band slope of our model is consistent with observations from GMRT and VLA telescopes. It also reproduces WISE far-IR data as well as observations from the WHT and LIV telescopes. The X-ray data observed by {\sl Swift} correspond to the high-energy tail of the synchrotron emission, whereas the Fermi-LAT $\gamma$-ray data are well described by the SSC emission.

The best fit shown in this figure is obtained assuming a redshift of $z=0.18$. Acceptable fits can also be achieved for redshifts in the $0.2$–$0.4$ range. For higher values, agreement with the data progressively deteriorates. In particular, redshifts $z\gtrsim 1$ are disfavored, as reproducing the observed $\gamma$-ray spectra would require invoking unusually large Doppler boosting factors or very hard particle injection spectra. While not excluded, such conditions depart from those commonly inferred in standard blazar emission models and are less naturally supported by the available multiwavelength constraints.

\begin{figure}[!t]
\centering
\includegraphics[width=1\linewidth]{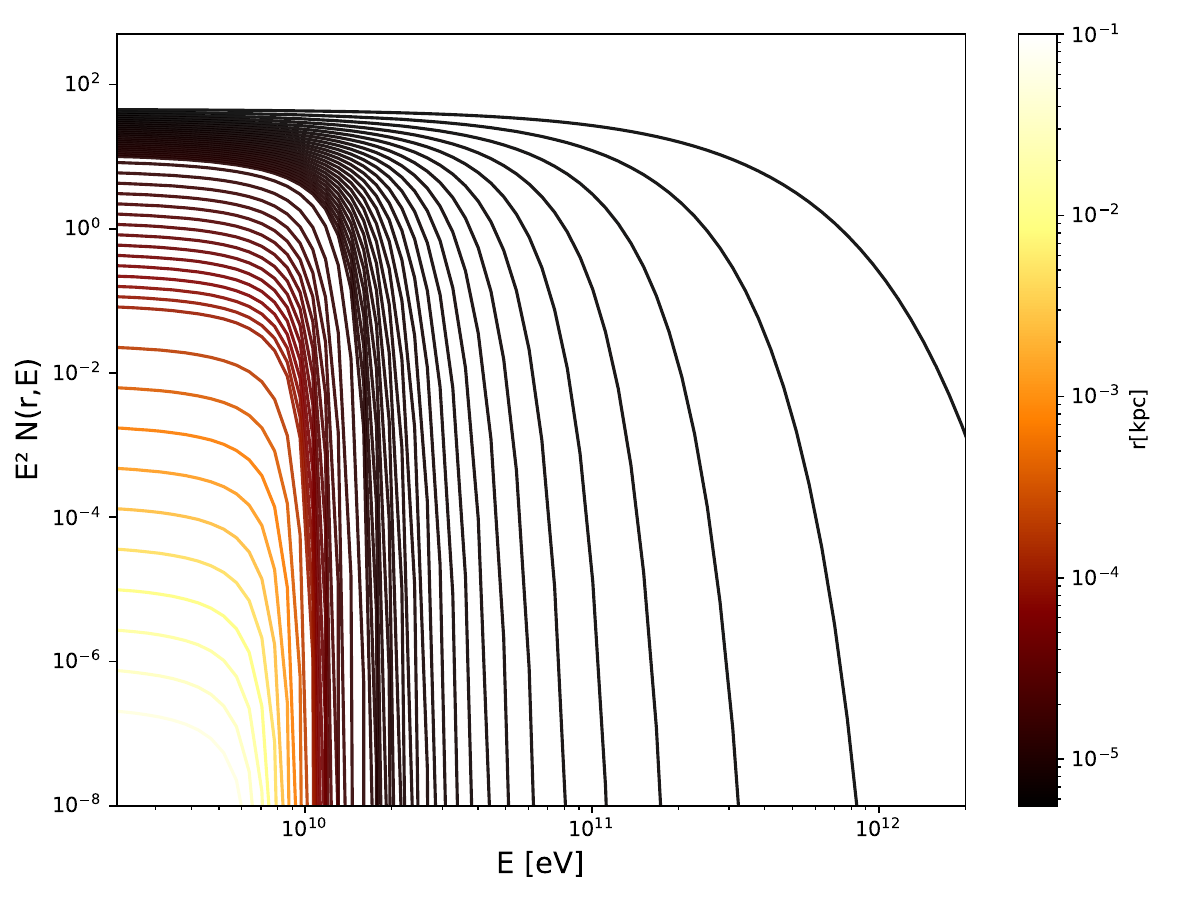}
\caption{Electron distribution $N_e(r,E)$ as function of energy $E$ for different distances $r$ along the jet. The parameters correspond to those listed in Table \ref{tabla}.}
\label{fig:NevsE}
\end{figure}

Finally, in Fig. \ref{fig:NevsE}, we show the corresponding electron energy distribution at different distances along the jet to illustrate the impact of the radiative cooling.

\begin{table} 
\centering
\caption{Input and derived parameters.}
\begin{tabular}{lcc}
\hline\hline\noalign{\smallskip}
\ \ Parameter & \ \ \ \  Description & \ \ \ \  Value  \\
\multicolumn{3}{c}{\textit{\rm}} \\

\hline\noalign{\smallskip}
  \ \ $L_{\rm j} \ ({\rm erg \  s^{-1}})$& jet luminosity & $10^{45}$\\  
  \ \ $d_L \ (\rm Mpc)$ & distance to the blazar & $878.3$  \\
      \ \ $z \ $ & redshift & $0.18$ \\    
  \ \ $M_{\rm BH} \ (M_{\odot})$ & black hole mass & $10^8$ \\
  \ \ $\Gamma \ $ & bulk Lorentz factor & $10$ \\    
      \ \ $i_{\rm j}$ & angle with the line of sight & $0.1$ \\
  \ \ $\theta_{\rm j} \approx 1/\Gamma$ & half opening angle & $0.1$ \\
\ \ $E_{\rm max} \ (\rm GeV)$ & maximum energy & 192.7 \\   
   \ \ $q_{\rm rel}$& ratio $L_{\rm e}/L_{\rm {\rm k}}$ & $0.1$ \\
 \ \ $\epsilon_{\rm B}$ & magnetic field energy fraction & $0.007$ \\
 \ \ $B_{\rm onset} \ (\rm G)$ & magnetic field at $r_{\rm onset}$ &  $0.22$ \\
   \ \ $\alpha$ & injection index &  $2$\\
      \ \ $r_{\rm onset} \ (\rm r_g)$ & onset position &  $1150$ \\
  \ \ $\Delta r \ (\rm pc)$ & cone extension & 100  \\ 
  \ \ $\gamma_{\rm min}^e \ $ & $e$ minimum Lorentz factor & $3.66\times 10^{3}$  \\    

  \hline
\label{tabla}
       \end{tabular} 
\end{table} 

\section{Discussion}\label{discussion}

We developed an extended conical jet model to reproduce the available multiwavelength data of J141046.00+740511.2, including radio, infrared, optical, X-ray, and $\gamma$-ray observations. We find that the model is in good agreement with the observed emission from radio to gamma rays. This provides a more consistent explanation than that offered by standard one-zone models, which does not account for the radio emission mainly due to the effect of SSA, as discussed above. This also shows the importance of considering jet geometry in the context of blazars, as its extended regions can significantly affect broadband emission. In particular, not only is the synchrotron emission no longer self-absorbed due to contributions from the farther regions of the jet, but the overall fit to the broadband data is also improved. Moreover, taking radio data into consideration constrains the parameters used to fit the higher-energy emission, thus limiting parameter degeneracy. The Swift/UVOT data are not contemporaneous with radio observations. We included them in our comparison, but the fact that our model overestimates them does not indicate a real tension.

The synchrotron peak for our best fit is $E_{\rm peak} \approx 2.3 \ \rm eV$, consistent with J141046.00+740511.2 being an IBL. 
This difference with respect to previous HBL classifications is likely due to the contribution from extended regions of the jet, which increases the low-frequency emission and results in a softer spectrum.

With our extended-jet SSC model, we obtain a minimum Lorentz factor in the $\gamma_{\rm min} \sim 10^{3-4}$ range, in agreement with the values reported in previous work (e.g., \citealt{Bonnoli15,Sciaccaluga2022}). The inferred magnetic field is well below equipartition, similar to the values commonly found in single-zone SSC fits for IBLs. Alternative frameworks that partially alleviate the requirement of such low magnetization include the spine–layer jet geometry \citep{Tavecchio2016} and lepto-hadronic scenarios, where the high-energy component can be ascribed to proton synchrotron; the latter will be explored in an extension of the present model.

From our modeled SED, we obtain a radio-loudness parameter of $\log(R)=0.9$. In the context of our discussion on whether the $R$ parameter accurately reflects the radio-loudness of blazars, this model shows that even for a high $R$ value (e.g., $\log(R)\simeq1$; see \citealt{Cao2019}, \citealt{Marchesini2023}, \citealt{Ulgiati2024}), the blazar emits significantly lower luminosity ($\rm L_{5GHz} = 1.97 \times10^{39} \rm erg/s$) in the radio band. This is due to the dominance of adiabatic cooling at distances far from the BH and the decline of the magnetic field along the jet, which together decrease the total radiative output of the electrons.

Following the approach of \citet{Mantovani2015}, we compared the radio flux density with the integrated $\gamma$-ray flux. In the absence of measurements at $15$ GHz, we estimated this value by extrapolating the slope of the radio emission measured at $3$ GHz with VLA. For WISE J141046.00+740511.2, we obtain a flux density of $3.97 \times 10^{-4}$ Jy at $15$ GHz and an integrated Fermi flux of $2.04\times 10^{-9}$ ph cm$^{-2}$ s$^{-1}$. These values place the source in the bottom-left region of the correlation plot (Fig. 4 from \citealt{Mantovani2015}), consistent with a very weak blazar.

Finally, it is important to emphasize that neither our model nor the parameters employed to reproduce the SED of the source require any extraordinary assumptions. The scenario we adopted corresponds to an extended jet that includes convective transport, and the physical parameters involved remain well within the typical ranges reported for BL Lac objects. This highlights that the emission features can be accounted for without invoking extreme conditions, but rather through a physically motivated framework consistent with the established phenomenology of blazar jets. Additional observations are required to better constrain the parameter space as well as to determine the redshift, which is unknown for this source. In future work we will apply the same model to other RWBL candidates and explore alternative scenarios such as structured jets and lepto-hadronic solutions.

\begin{acknowledgements}
      We thank the referee for their careful reading of the manuscript and for their constructive comments, which have helped improve the clarity and quality of this work. This work was partially supported by ANPCyT and Universidad Nacional de Mar del Plata through grants PICT 2021-GRF-T1-00725 and 80020240500217MP, respectively. F.L.V. acknowledges support from CONICET (PIP 2021-0554).
\end{acknowledgements}

  \bibliographystyle{aa} 
   \bibliography{biblio.bib}

\begin{appendix}
\section{Nonthermal processes}\label{nt-processes}

\subsection{Synchrotron radiation}

The spectral power per unit energy radiated by an electron can be expressed in the jet reference frame as \citep{BlumenthalGould1970}

\begin{equation}
P_{\rm syn}(E_{\rm ph},E) = \frac{\sqrt{2}e^{3}B}{m_{e}c^{2}h} \frac{E_{\rm ph}}{E_{\rm cr}} \int_{\frac{E_{\rm ph}}{E_{\rm cr}}}^{\infty} d\zeta K_{\frac{5}{3}}(\zeta)
\end{equation}

\noindent Here, $K_{\frac{5}{3}}(\zeta)$ is the modified Bessel function of order $5/3$ and
\begin{equation}
E_{\rm cr} = \frac{\sqrt{6}heB}{4\pi m_{e}c} \left( \frac{E}{m_{e}c^{2}} \right)^{2} \ .
\end{equation}

The power per unit energy, per unit volume emitted by the electrons through synchrotron radiation, can be calculated as:

\begin{equation} \label{synemis}
\varepsilon_{\rm syn}(r, E_{\rm ph}) = \left(\frac{1 - e^{-\tau_{\rm SSA}(r,E_{\rm ph})}}{\tau_{\rm SSA}(r,E_{\rm ph})} \right) \int_{m_{e} c^{2}}^{\infty} dE 4 \pi P_{\rm syn} N_{e}(r, E),
\end{equation}

\noindent and thus, the emissivity can be expressed as

\begin{equation}
    Q_{\rm syn}(r,E_{\rm ph}) = \frac{\varepsilon_{\rm syn}(r,E_{\rm ph})}{E_{\rm ph}} \ .
\end{equation}

The factor $\left(\frac{1 - e^{-\tau_{\rm SSA}(r, E_{\rm ph})}}{\tau_{\rm SSA}(r, E_{\rm ph})} \right)$ in Eq. (\ref{synemis}) describes the effect of synchrotron self-absorption (SSA), which refers to the possibility of synchrotron-emitted photons being reabsorbed by the electrons themselves. The optical depth $\tau_{\rm SSA}$ associated with this can be written as
\begin{equation}
\tau_{\rm SSA}(r, E_{\rm ph}) \approx R(r) \alpha_{\rm SSA}(r, E_{\rm ph})  
\end{equation}

\noindent where $R(r) = r \, \tan \theta_{\rm j}$ is the radius of the jet at position $r$. The self-absorption coefficient, $\alpha_{\rm SSA}$, can be expressed as \citep{RybickiLightman1986}
\begin{equation}
\alpha_{\rm SSA} = -\frac{h^{3}c^{2}}{8\pi E_{\rm ph}^{2}} \int_{m_{e}c^{2}}^{\infty} dE E^{2} P_{\rm syn}(E_{\rm ph},E) \frac{\partial}{\partial E} \left[\frac{N_{e}(r, E)}{E^{2}}\right].
\end{equation}
This effect becomes relevant at low photon energies. In one-zone models, it can even suppress the emission in the radio band. This is not the case for extended jet models like the one we developed here, mainly due to the contributions from positions farther away from the BH.
In our model, we evaluated these expressions at each position $r$ along the jet, using the magnetic field $B(r)$ and the electron distribution $N_e(r, E)$ obtained from the solution of the transport equation.

\subsection{Inverse Compton radiation}

For an isotropic distribution of photons with density \( v d\epsilon \) and energy in the interval \( d\epsilon \), which scatter through interaction with an isotropic distribution of electrons, assuming the Thomson limit, and that in the electron's rest frame the energy transfer can be neglected, the net power lost by an electron in the process, and thus converted into radiation, is \citep{RybickiLightman1986}

\begin{equation} \label{eqn:pic} 
P_{\rm IC} = \frac{4}{3} \sigma_{T} c \gamma^{2} \beta^{2} U_{\rm ph}
\end{equation}

where 

\begin{equation} 
U_{\rm ph} \equiv \int \epsilon v d\epsilon \ .
\end{equation}

From Eq. (\ref{eqn:pic}), the total radiated power per unit volume for a medium of relativistic electrons can be calculated as:

\begin{equation} \label{eqn:pictot}
P_{\rm tot,IC}^{\rm relativistic} = \int P_{\rm IC} N_e(\gamma) d\gamma \ ,
\end{equation}

\noindent where \( N_e(\gamma) d\gamma \) is the number of electrons per unit volume with \( \gamma \) in the interval \([ \gamma, \gamma + d\gamma ]\).

The expression in Eq. (\ref{eqn:pic}) is valid in the Thomson regime, where $\gamma E_{\rm ph} \ll m_e c^2$. Thus, in order to calculate the emissivity due to IC scattering in the context of high energies, as is the source considered here, it is important to consider the full Klein–Nishina cross-section. The expression adopted is:
\begin{equation} \label{eqn:QIC}
  Q_{\gamma,\rm IC}(r, E_{\gamma}) = \frac{r_{e}^{2}c}{2} \int_{E_{\rm ph}^{(min)}}^{E_{\gamma}} dE_{\rm ph} \frac{n_{\rm ph}(r,E_{\rm ph})}{E_{\rm ph}} \int_{E_{\rm min}}^{E_{\rm max}} dE \frac{N_{e}(r,E)}{\gamma_{e}^{2}} F(q),
\end{equation}

\noindent Here, $r_{e}$ is the classical radius of the electron, $E_{\rm ph}$ is the energy of the target photons, and $E_{\gamma}$ is the energy of scattered photons. The factor $F(q)$ is given by:

\begin{equation} 
F(q) = q\ln q + (1+2q)(1-q)+\frac{1}{2}(1-q)\frac{(q\Gamma_{e})^{2}}{1+\Gamma_{e}},
\end{equation}
\noindent where $\Gamma_{e}=4E_{\rm ph}E/(m_{e}^{2}c^{4})$, and

\begin{equation} 
q = \frac{E_{\gamma}}{\Gamma_{e}E_{\rm ph} \left(1-\frac{E_{\gamma}}{E_{\rm ph}}\right)}.
\end{equation}

\noindent The integration limits for the electron energies on Eq. (\ref{eqn:QIC}) are:

\begin{equation}
\begin{aligned}
 E_{\rm min} &= \frac{E_{\gamma}}{2} + \frac{m_{e}c^{2}}{2} \sqrt{\frac{E_{\gamma}}{2E_{\rm ph}}+\frac{E_{\gamma}^{2}}{2m_{e}^{2}c^{4}}} ,\\
E_{\rm max} &= \frac{E_{\gamma}}{1-\frac{E_{\gamma}}{E_{\rm ph}}} .
\end{aligned}
\end{equation}

Since the photon field itself is produced by synchrotron radiation from the same electrons that undergo IC cooling, the electron distribution $N_{e}(r,E)$ and the radiation field are intrinsically coupled. To ensure a self-consistent solution, we adopted an iterative procedure, similar to those commonly used when solving coupled kinetic equations for particles and photons in scenarios where the photon field is generated by the particle population itself \citep{Vurm2009,Schlickeiser2010,Vieyro2012,Boetcher2013}. In our model, synchrotron losses dominate over most of the jet, although IC cooling becomes relevant in the most compact regions. In this way, our approach remains valid beyond the purely linear approximation commonly adopted in SSC models, while naturally recovering it in the regime where the feedback of the photon field on the electron distribution is weak.

The numerical scheme proceeds as follows. First, the transport equation for the electron distribution is solved including only synchrotron and adiabatic losses. From this solution, we computed the synchrotron emissivity and the corresponding photon density. Using this photon field, we then calculated the IC emissivity and the associated cooling rate. The electron transport equation is subsequently solved again with the updated radiative losses. This procedure is repeated until convergence of the electron distribution is achieved. In practice, convergence is typically reached after a few iterations, usually fewer than five. 

\begin{figure}[!t]
\centering
\includegraphics[width=1\linewidth]{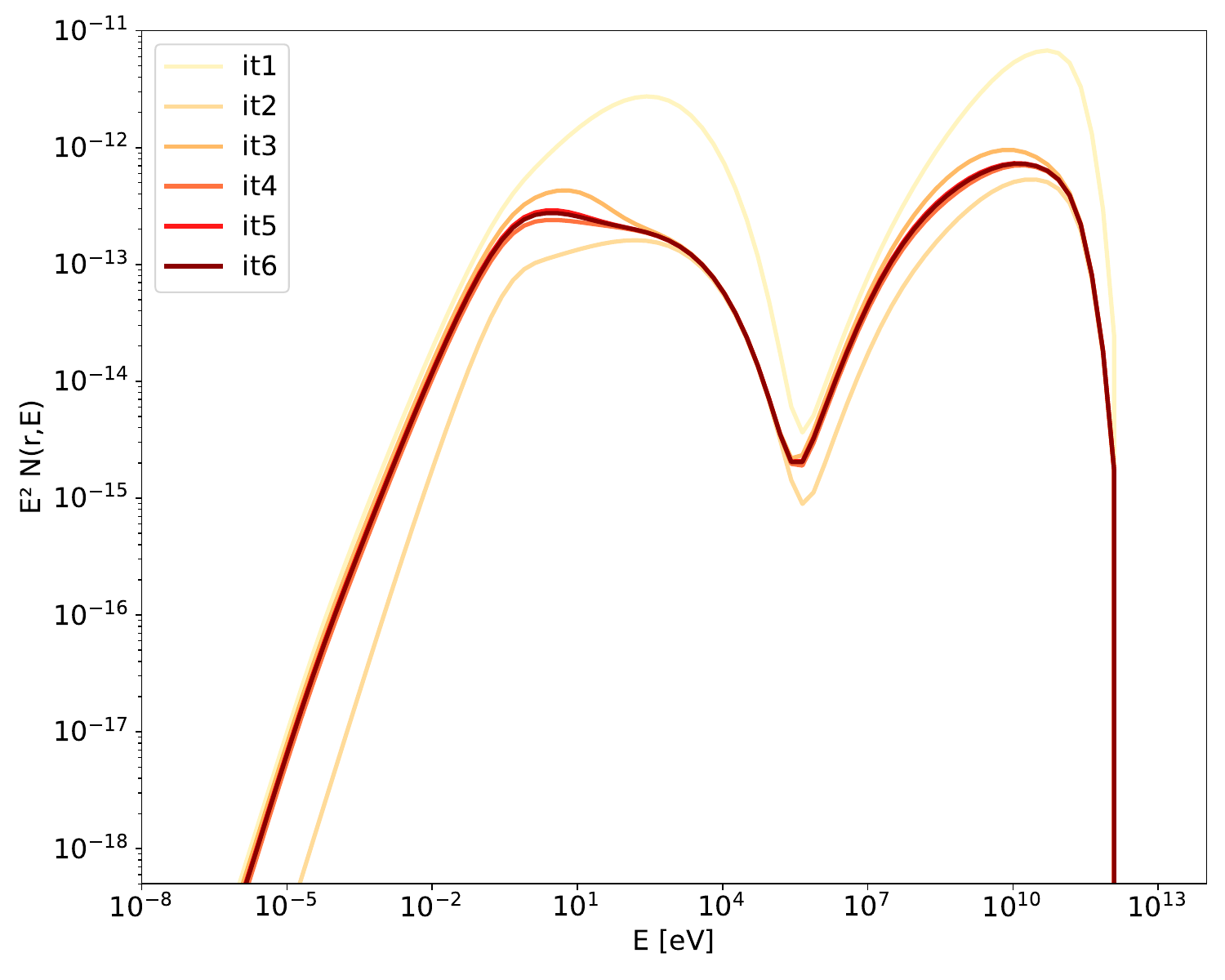}
\caption{Fluxes corresponding to different SSC iterations.}
\label{fig:Qphit}
\end{figure}

In Fig. \ref{fig:Qphit}, we show the evolution of $Q_{\gamma,\rm IC}(r, E_{\gamma})$ over several iterations. The electron distribution is only weakly modified between successive iterations, with changes limited to the maximum electron energy. This behavior indicates that the system operates close to the linear synchrotron-dominated regime, where $U_{\rm B} \gtrsim U_{\rm syn}$ and IC losses provide only a subdominant contribution to the total cooling rate.

\end{appendix}

\end{document}